\documentclass[newstyle,twocolumn,proceedings]{rmaa}
\usepackage{rmaacite}
\renewcommand{\P}[1]{%
\ifnum#1=1\hbox{OW~168--326E}\fi
\ifnum#1=2\hbox{OW~167--317}\fi
\ifnum#1=3\hbox{OW~163--317}\fi
\ifnum#1=5\hbox{OW~158--323}\fi
\ifnum#1=0\hbox{OW~171--334}\fi}

\def\ea{{\it et al}~}

\title{Magnetized Astrophysical Outflows: \\
Cradle to Grave, Source to Effect}
\author{A. Frank, T. Gardiner\altaffilmark{1} and T. Lery\altaffilmark{2}}
\altaffiltext{1}{Department of Physics and Astronomy, University
    of Rochester, USA\\}
\altaffiltext{2}{DIAS, School of Cosmic Physics, 5 Merrion Square,
    Dublin 2 Ireland\\}
\fulladdresses{
\item Adam Frank: Department of Physics and Astronomy, University
    of Rochester, Rochester, NY 14627, USA(afrank@pas.rochester.edu).}
\shortauthor{Frank, Gardiner \& Lery}
\shorttitle{Magnetized Jets}

\abstract{

We discuss the propagation of radiative MHD jets and outflows
focusing on outflows driven by magneto-centrifugal rotators. Our goal is
to link the properties of the jets with the physics of the sources
which produce them. We find that density and magnetic field
stratification (with radius) in jets from magnetized rotators
leads to new behavior including the development of a dense inner jet
core and a low density collar.  We also report on more general studies
of ambipolar diffusion and field geometry in pulsed jets. Finally we
describe a new work designed to study the effects of magnetized winds
on circumstellar environments appropriate to YSOs and PNe.

}
\listofauthors{Frank, A., Gardiner, T. A., \& Lery, T.}
\indexauthor{Frank, A.}
\indexauthor{Gardiner, T. A.}
\indexauthor{Lery, T.}
\begin{document}

%% This command is necessary to typeset the title, abstract, etc.
\maketitle

%%
%% And here starts the text....
%%
\section{Introduction}
\label{sec:intro}

Hypersonic collimated flows (jets) are a ubiquitous phenomena in
astrophysics occurring in wide variety of environments including
young stars (YSOs) \cite{Reipurth97} and evolved stars (PNe)
\cite{SokLiv94}.  In almost all cases magnetic fields are expected
to play a key role in launching and collimating these outflows. In
particular, magneto-centrifugal processes associated with
magnetized, rotating stars \cite{bots99} and/or accretion disks
\cite{Pudritz91} are believed to lift material out of the
gravitational well and provide at least some confinement, shaping
the wind into a jet. There has been considerable effort in the study
of collimated outflows over the last two decades both in terms of jet
simulations and the study of magneto-centrifugal
launching. Unfortunately, direct links between the observable jets
(scales of $> 10^{15} ~cm$) and the collimation processes (scales of
$<10^{13} ~cm$) have yet to be established. Indeed, until recently the
majority of jet propagation simulations have been purely
hydrodynamic. Thus the role magnetic fields play in establishing the
properties and behavior of radiative jets (appropriate to YSOs and
PNe) remains an open issue. In addition, establishing links between
the near-field (close to the star) magneto-centrifugal processes and
the far-field jet behavior which can be readily observed must also be
established.

In this contribution we describe the MHD behavior in radiative jets
and outflows.  The goal of our studies is to articulate how magnetic
fields effect radiative jet/outflow behavior and to provide links
between magneto-centrifugal processes occurring at the source with
observable properties of the jets and outflows.

\section{Jet Structure and the Problem of Initial Conditions}
\label{sec:constraints}

Magnetic fields imposed on the flow during the launching process may
bleed out the beam by {\it ambipolar diffusion} (where ions and the
field slip past neutrals. \scite{Frankea99} have shown that the
ambipolar diffusion timescale in YSO jets can be written as
\begin{displaymath}
t_{ad} \approx  10^4 \left( { n_j \over 10^3 cc} \right) \left( {
R_j \over 10^{15} cm} \right)^2 \left( { 10^4 K \over T_j} \right)
Q(\beta)
\end{displaymath}
\noindent where $\beta$ is the ratio of gas to magnetic pressure,
$Q(\beta) = \beta / (\beta + 1)$, and $n_j$ refers to the neutral
fraction in the jet beam.  Since the largest scale YSO jets have
dynamical times of order a few $10^4$ to $10^5 ~\textrm{y}$ 
\cite{Reipurth97} our result shows that fields will remain in the
beam for much of the jet's lifetime. The characteristic ambipolar
diffusion timescale of $10^4 ~\textrm{y}$ is also suggestive. It's
approximate equivalence with the age of parsec-scale jets may indicate
that these structures remain intact as long as the collimating fields
remain in the beam.  This issue needs further study.

Establishing initial equilibria for MHD jet simulations is
non-trivial.  This is not the case for hydrodynamic jets where the
required force balance across the jet and ambient medium interface
allows for the use of so-called {\it top-hat} profiles (i.e. the hydro
variables are constant across the jet cross section).  Such
distributions may not be tenable in MHD jet studies. The difficulty
can be seen by decomposing the Lorentz force into a tension term and a
pressure term. ${\bf F_l} \propto -\nabla {\bf B}^2 + 2({\bf B} \cdot
\nabla) {\bf B}$. In a steady, cylindrically collimated jet only
$B_\phi$ and $B_z$ components of the field are possible.  Jets with
purely longitudinal fields, $B=B_z$, can be easily set in pressure
balance with the environment and top-hat profiles may be
used. Toroidal or helical field geometries require more complicated
initial conditions unless the field is assumed to take on a force
free configuration. If the field is not force free, MHD jets must have
variable distributions of gas pressure and, perhaps, other variables
in order to balance the {\it hoop stresses} associated with the
tension force.

Faced with the problem of initial conditions researchers studying
radiative MHD jets have, in general adopted one of two strategies:
(i) use force-free fields \cite{Cerqueira98,Cenquria99};
(ii) use ad-hoc gas pressure and magnetic field distributions
configured to be in initial force balance
\cite{Frankea99,StoneHardee00,OsulRay00}. The results of these
studies for both steady and time-variable (pulsing) jets 
reveal a number generic features.
\begin{itemize}
\item{Jets with purely longitudinal geometries do not
show propagation characteristics which differ significantly from the
hydrodynamic case \cite{Cerqueira98,Gardinerea00}. The presence of
poloidal fields does however allow for the possibility of field
reversals at the head of the jet which are likely to be unstable to
reconnection \cite{Gardinerea00}. The transfer of magnetic energy
into thermal energy at reconnection sites may alter the emitted
spectrum away from that associated with shocks.  Given the importance
of shock diagnostics for interpretation of spectra the additional
source of excitation provided by reconnection requires further study.}
\item{Jets with a toroidal field component will be subject to
strong hoop stresses especially in the region between the jet and bow
shocks \cite{Clarke86,Frankea98}. In 2.5-D axisymmetric calculations
pinch forces associated with the toroidal fields lead to the
development of streamlined, high field {\it nose-cones}. Recent
simulations have shown such nose-cone may be unstable in 3-D (Dal
Pino, these proceedings). While this result still needs further study,
it may be correct for jets with top-hat density profiles.  As we
will show in the next section, more realistic initial conditions for
jets from magneto-centrifugal rotators may not have unstable nose-cones.}
\end{itemize}

While the studies discussed above have yielded significant progress in our
understanding of heavy, radiative MHD jet behavior, the initial
conditions are still unconnected to the processes believed to create
the jets.  We turn to this issue in the next section.

\section{Jets From Keplerian Rotators}

Recently we have simulated jets whose initial conditions jets are
taken directly from a (simplified) model of the magneto-centrifugal
launching/collimation process.  The model, known as the Given Geometry
Method (GGM) allows asymptotic MHD jet equilibria to be linked to the
properties of a rotating source
\cite{Leryea99,LeryFrank99}. The GGM assumes a
time-independent, axisymmetric flow.  It further simplifies the
problem of magneto-centrifugal launching/collimation by assuming that
the nested magnetic flux surfaces defining the flow possess a shape
which is known a priori inside the fast critical surface (the locus of
points beyond which the flow is kinetic energy dominated).  The flux
surfaces are assumed to be conical and, as an additional
simplification, an equilibrium across the surfaces is assumed at the
Alfv\'en point.  The final asymptotic solution for the collimated jet
is solved by assuming pressure balance with the ambient medium.

In terms of dynamics, The most important properties of the jets
derived via the GGM are the radial variations in the density $\rho(r)$
and the toroidal component of the magnetic field $B_\phi(r)$. In
particular when the source is composed of a rapidly rotating disk
truncated some distance from a rigidly rotating star, the emitted jets
can have strong density stratifications, ie. a high density
axial {\em core} surrounded by a lower density {\em
collar}. The strongest toroidal field is located at the boundary of
the core and collar creating a magnetically confined jet-within-a-jet
structure. Note that the bulk of the jet's momentum resides in the
core. Hence we expect this portion of the beam to penetrate more
easily into the ambient medium during the jet's propagation while the
collar will be more strongly decelerated.

Figure 1 (top) shows a frame from a jet simulation with a high core to
collar density ratio ($\sim 100$, see \scite{Frankea2000}).  As
expected the jet core propagates faster through the grid then the
outer collar.  In these simulations, (which do not include radiative
losses), the collar is diverted by shocks at the initial head of the
jet. The collar peels away as the jet propagates down the grid.  A
distinct nose-cone forms at the head of the jet core due to its strong
confining toroidal magnetic field.  Nose-cones associated with this
high density core may not be unstable. Also of particular interest are the
apparent pinch mode ($m=0$) instabilities seen in the core.  A
stability analysis of the initial conditions effectively predicts the
wavelength of these modes ($\lambda = .3 R_j$) and suggests these they
are current driven rather hydrodynamic Kelvin-Helmholtz instabilities.

Since many jets are seen to be episodic we have can also use the GGM
method to derived a sequence of jet solutions appropriate to a time
variable flow.  Figure 1 (bottom) shows the density in a radiative
pulsed jet simulation \cite{GardinerFrankLery2001} from a Keplerian
rotator.  Here again we see the core/collar structure which now
persists in the internal working surfaces. The collar is not seen to
peel away from the core in these models, a result which may be due to
the pulsing or to the lower core to collar density ratio obtained from
the GGM model in this case. The eventual goal of these studies will be
to model jets emitted by sources undergoing some form of FU Ori type
outburst and hence connect the properties of variable jets to
theoretical models of variations at the source.

\begin{figure}
\includegraphics[width=\columnwidth]{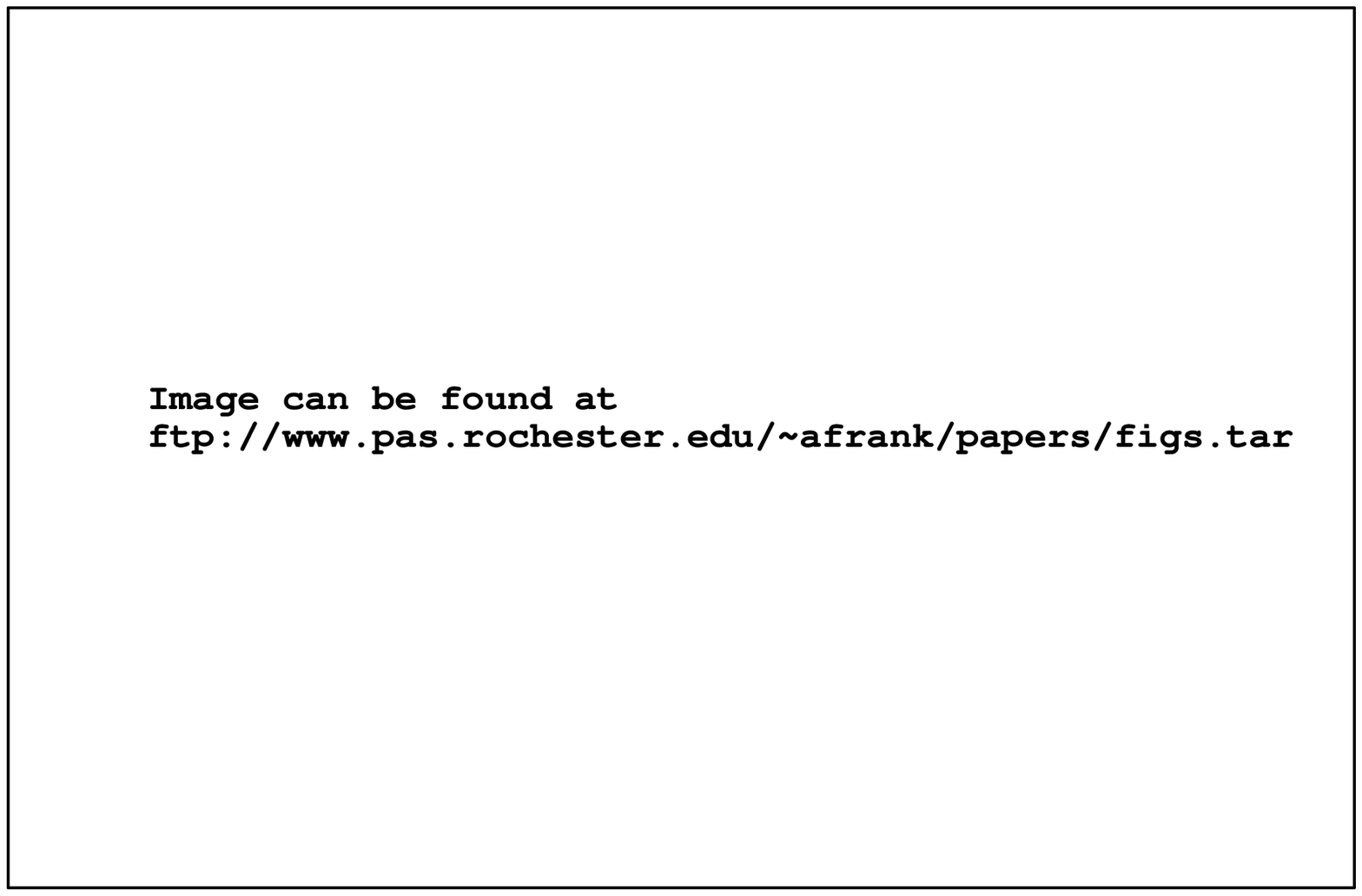}
\caption{Top: Grey-scale map of the density in steady jet calculation
from a disk+star rotator (see \pcite{Frankea2000} for details). Bottom:
Pulsed jet simulation. Initial conditions derived from a sequence of
GGM Keplerian rotator models \cite{GardinerFrankLery2001}.}
\end{figure}

\section{The Magnetic Geometry of Pulsed Jets}
The most general magnetic geometry for cylindrically collimated jets
is a helical field.  Indeed, this is the geometry which is to be
expected from magneto-centrifugal launching. Since many jets are
observed to episodic, it is of interest to understand how variations
in the flow at the source will effect the subsequent downstream field
geometry of the jets. In \scite{GardinerFrank2001a} a simple,
universal relation was derived for the field configuration which would
evolve in a helically magnetized, pulsed jet.

Consideration of the governing flow equations reveals that the
toroidal field component $B_{\phi}$ will evolve in manner similar to
the gas density. Thus, downstream compressions produced by internal
working surfaces will also lead to compression in the toroidal
field. Likewise, regions of gas rarefactions will also become regions
of weak toroidal field.  \scite{GardinerFrank2001a} derived an
expression for the ratio of the toroidal to poloidal field components
at a time $t$ based on a Burger's equation analysis,
\begin{displaymath}
\frac{B_{\phi}}{B_z}=\frac{B_{\phi,0}}{B_z}\left(\frac{1}{1-k(t-t_o)}\right)
\end{displaymath}
where $t_o$ is the time when the gas parcel was ejected and $k$ is the
ratio of the derivative of the jet ejection velocity at $t_o$ to the
value of the velocity at that time $k=v_j(t_o)'/v_j(t_o)$.

The relation above shows that pulsed helical jets will develop
alternating regions of toroidal and poloidal domination in the
beam.  High density knots associated with the internal working
surfaces will be toroidally dominated and the inter-knot rarefied
regions will be poloidally dominated.  Figure 2 shows a plot of
$B_\phi$ at two different times from a pulsed jet simulation.  The
development of strong field regions is clearly seen as the pulses
steepen into internal shocks.

\begin{figure}
\includegraphics[width=\columnwidth]{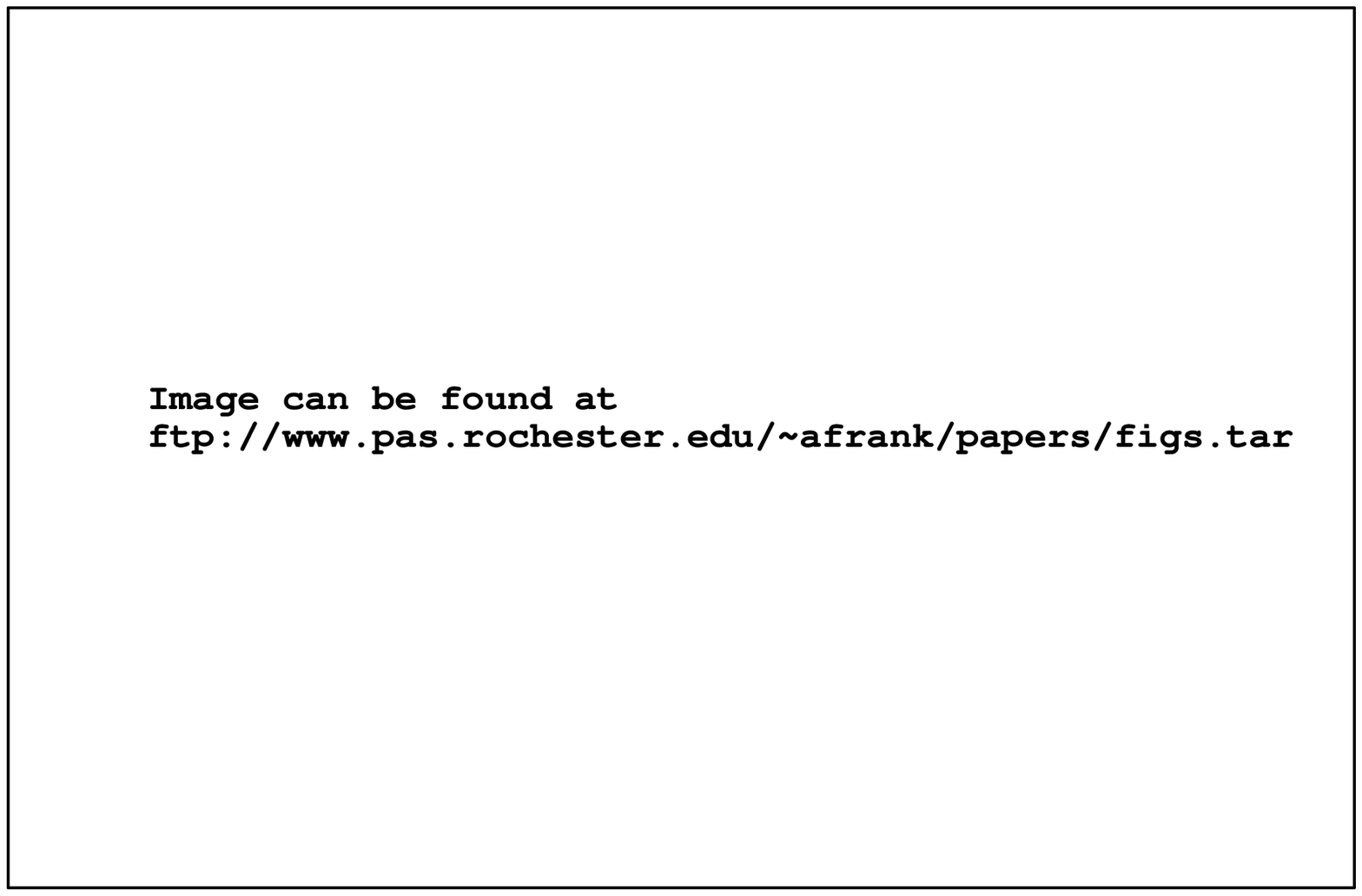}
\caption{$Log_{10}(B_\phi/B_{\phi0})$ at two different times in a
pulsed radiative MHD jet simulation with an initial helical field. For
details see \scite{GardinerFrank2001a}.}
\end{figure}

\section{Magnetized Wind Blown Bubbles:YSO and PNe} 
In previous sections we discussed the behavior of fully
collimated jets.  In both YSOs and PNe, however, the outflows may
include a significant {\it wide-angle} magnetized wind component.
Numerous studies have shown that the pure hydrodynamical interaction
of a wide-angle wind with an aspherical ambient density distribution can 
lead to highly collimated bipolar flows
\cite{Ickeea92,DelamarterFrankHartmann}.  Since the environment
surrounding both YSOs and PNe is expected to include an aspherically
shaped density distribution, the additional effect of magnetic fields
in shaping the outflows is of great interest.

A number of authors have explored a model whereby a weak toroidally
magnetized wind is collimated by hoop stresses after passage through a
wind shock \cite{ChevalierLuo,GarciaSegura99}.  This so-called {\it
magnetized wind bubble} (MWB) model shows some promise in that highly
collimated outflows can be obtained even when purely hydrodynamic
flows would only yield rather wide bipolar lobes.  Precession of the 
sources' magnetic axis also allows for
point-symmetry, (observed in many PNe), to be imposed on the flows
\cite{Sa00}.

Recently however \cite{GardinerFrank2001b} have shown that the MWB
fails to account for collimation which would occur {\it before} the wind is
shocked.  The toroidal field in the wind produces an unbalanced
collimating force which can redirect the wind as it freely
expands. Thus a proper treatment of this model must include the full
history of the wind starting close to the source.  It is also noteworthy 
that the MWB model is consistent
with a wind accelerated by some means other than magneto-centrifugal
launching.

Figure 3 shows a simulation of a magnetized wind blown bubble
appropriate to YSOs \cite{GardinerFrankHartmann}. In this model a {\it
pre-collimated} wind with a pole to equator density ratio of $10$ is
driven into a collapsing sheet-like environment.  The initial
conditions capture many of the salient properties of
magneto-centrifugal flows such as the X-wind \cite{Shuea94,Matzner99}.  The
figure shows a strongly collimated flow, including a jet, forming along
the axis.  Note that though the flow begins with a strong axial
condensation it is not visible until after
the wind passes through the shock.  As with previous studies of the MWB 
model this occurs due to
strengthening of hoop stresses associated with the toroidal field
after the wind has been shocked.

\begin{figure}
  \includegraphics[width=\columnwidth]{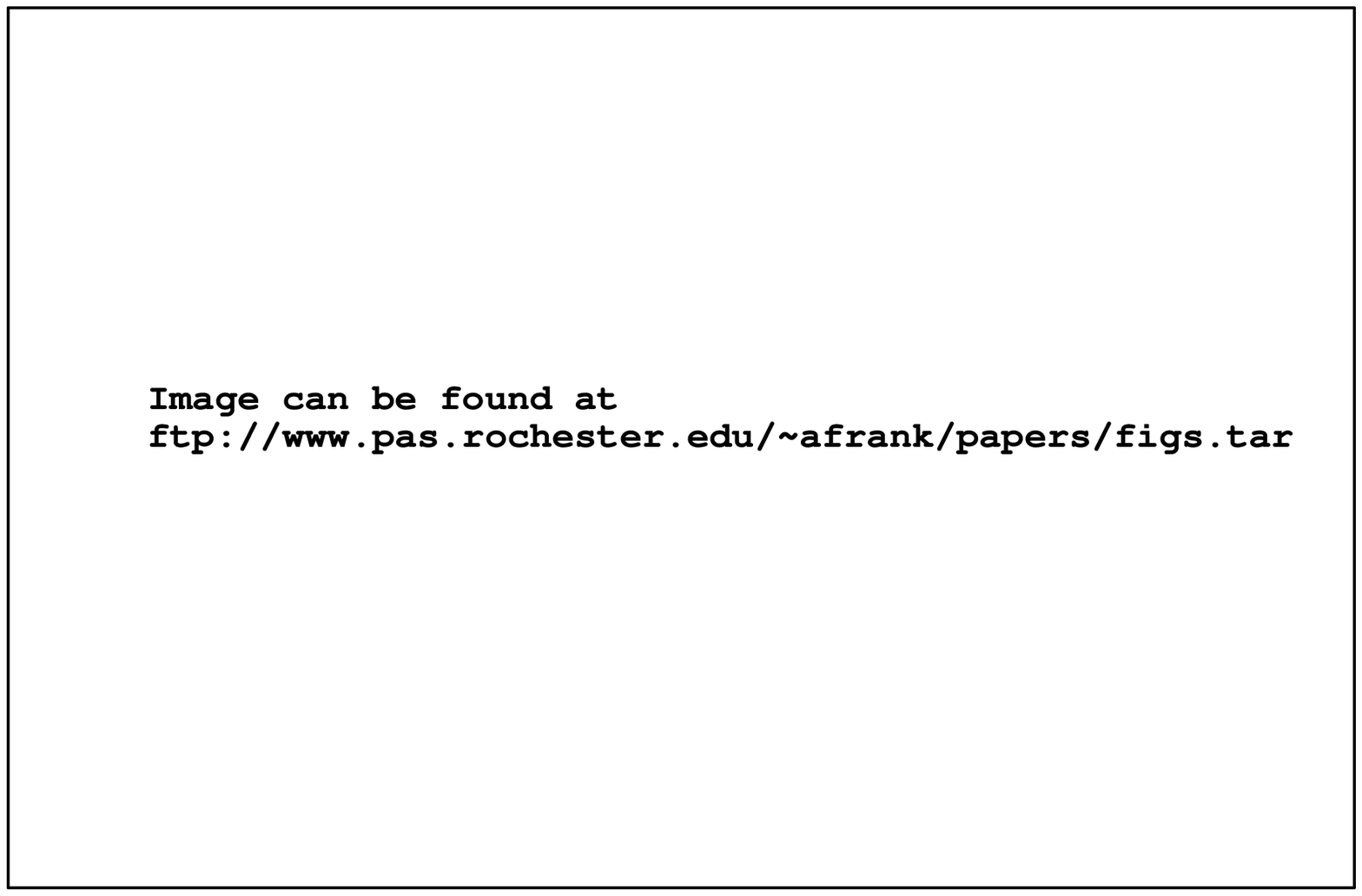}
  \caption{Simulation of MWB model in a collapsing environment appropriate
to YSOs.  The simulation begins with a wind with properties similar to
those from magneto-centrifugal flows.}
\end{figure}

\section{Conclusions}
Magnetic fields offer a theoretically attractive means for producing
and shaping much of the outflow behavior observed in both young and
evolved stars.  The obvious failing of magnetic models rests in
the unfortunate reality that the fields themselves are, in most cases,
unobservable with current techniques.  Thus the presence and effect of
the these important players on the stage of stellar evolution must be
inferred by other means. Currently it is not clear if YSO jets, the
structures most likely associated with stellar or disk magnetic
fields, show any behaviors, such as nose-cones, which can be clearly
linked to the presence of magnetic stresses.  It would be an ironic
situation indeed if the most favored model of jet launching and
collimation leaves us with no signature with which to confirm its
veracity.

It is still too early to know if simulation studies can provide proxy
links between outflows and magnetic fields. We believe, however, that
the use of initial conditions which come directly from MHD launching
and collimation models should provide the quickest route to a final
answer.

\acknowledgements 
We are very grateful to Eric Blackman and Bruce Balick for discussions
and to Tom Jones and Dongsu Ryu for their insights and the use of some
numerical tools.  This work was supported by NSF Grant AST-0978765,
NASA Grant NAG5-8428, and the University of Rochester Laboratory for
Laser Energetics.

%% When using the rmaacite package, the \bibitem command should be of
%% the format:
%%
%%             \bibitem[AUTHOR<YEAR>]{KEY}
%%
%% so that the \cite{KEY}, etc. commands will work properly.
%%
%% If you are doing the citations manually, then you can just use
%% `\bibitem{}' instead. This will give you a warning about
%% `multiply-defined labels' which you can safely ignore.
%%

\end{document}